\documentclass[prb,twocolumn,showpacs,superscriptaddress,floatfix]{revtex4}

\usepackage{graphicx,amsmath,}
\usepackage{bm}

\newcommand{\ij}{i\kern -0.08em j}

\begin{document}

\title{Possible implementation of adiabatic quantum algorithm with
superconducting flux qubits}
\author{M.~Grajcar}
\altaffiliation{Electronic address: grajcar@fmph.uniba.sk}
\affiliation{Department of Solid State Physics, Comenius University,
SK-84248 Bratislava, Slovakia}
\affiliation{%
Institute for Physical High Technology, P.O. Box 100239, D-07702
Jena, Germany}
\author{A.~Izmalkov}
\affiliation{%
Institute for Physical High Technology, P.O. Box 100239, D-07702
Jena, Germany}
\affiliation{%
Moscow Engineering Physics Institute (State University),
Kashirskoe sh.\ 31, 115409 Moscow, Russia}
\author{E.~Il'ichev}
\affiliation{%
Institute for Physical High Technology, P.O. Box 100239, D-07702
Jena, Germany}

\date{\today}

\begin{abstract}

We show that an $LC$ parametric transducer can be effectively used
to monitor an adiabatic evolution of the superconducting flux qubit.
We propose a new scheme to measure the qubit's state, which is a
quantum nondemolition measurement. The scheme can be easily extended
to a three-qubit system, and  allows the reading out of the qubits'
states while the system remains in the ground state. An
implementation of the adiabatic quantum algorithm MAXCUT for three
superconducting flux qubits is discussed.
\end{abstract}


\maketitle

\section{Introduction}

Ten years ago Peter Shor demonstrated theoretically\cite{Shor94}
that a quantum computer can factor large numbers much more
effectively than a classical one. This discovery started an enormous
effort to find a physical system which would be a suitable qubit,
the building block of a quantum computer. Qubits are effectively
two-level systems with controlled parameters. There are many systems
in physics which can play the role of a qubit. One of them is a
superconducting flux qubit which can be realized as a
superconducting loop with low inductance $L_q$ interrupted by three
Josephson junctions. Its properties have already been
analysed\cite{Mooij99,Orlando} and experimentally
verified.\cite{Wal00} Superconducting qubits have several advantages
over qubits based on microscopic systems: they are scalable and  can
be accessed more easily and controlled individually. Moreover,
aluminum technology, widely exploited for the preparation of
conventional silicon devices, can be used.

Recently, several groups  succeeded in demonstrating coherent
macroscopic tunneling and Rabi oscillations in superconducting
qubits. This can be considered as the first important step towards
quantum computer realization.\cite{macro,Chiorescu03,Rabi} Most of
them were time domain measurements, which are supposed to be
important for quantum computing, since the much effort has been made
in the direction of building a quantum computer based on a universal
set of gates. However, in order to run an algorithm on such a
universal quantum computer, quantum error corrections should be
implemented. For a solid state qubit the error rate is only slightly
below the threshold required for fault-tolerant computation. This
places tremendous requirements on the hardware:\cite{Oskin02} the
number of physical qubits should be larger than $10^4$ and
teleportation between each two qubits should be possible. On the
other hand, the new scheme of quantum computation based on adiabatic
quantum evolution, which has been proposed by Farhi et
al.,\cite{Farhi00} could solve tasks beyond the reach of present-day
classical computers for a very moderate number of qubits ($\gtrsim
30$). Very recently a scalable superconducting architecture for
adiabatic quantum computation  was proposed which requires
nearest-neighbor coupling only.\cite{Kaminsky04} Moreover, Aharonov
et al.\cite{Aharonov04} have shown that adiabatic quantum
computation is equivalent to standard quantum computation. From an
experimental point of view the adiabatic quantum algorithm MAXCUT
was demonstrated by an NMR technique on three qubit
systems.\cite{Steffen03} In this paper, we propose a specific
implementation for  adiabatic quantum computing with a set of
coupled superconducting flux qubits, which is possible to realize
with the present state of the art. We show that a parametric
transducer can be effectively used to read out the results of the
adiabatic evolution algorithm.

\section{Parametric transducer as a QND readout for adiabatic quantum
computation} 
\begin{figure}[tbp]
\centering \includegraphics[width=8cm]{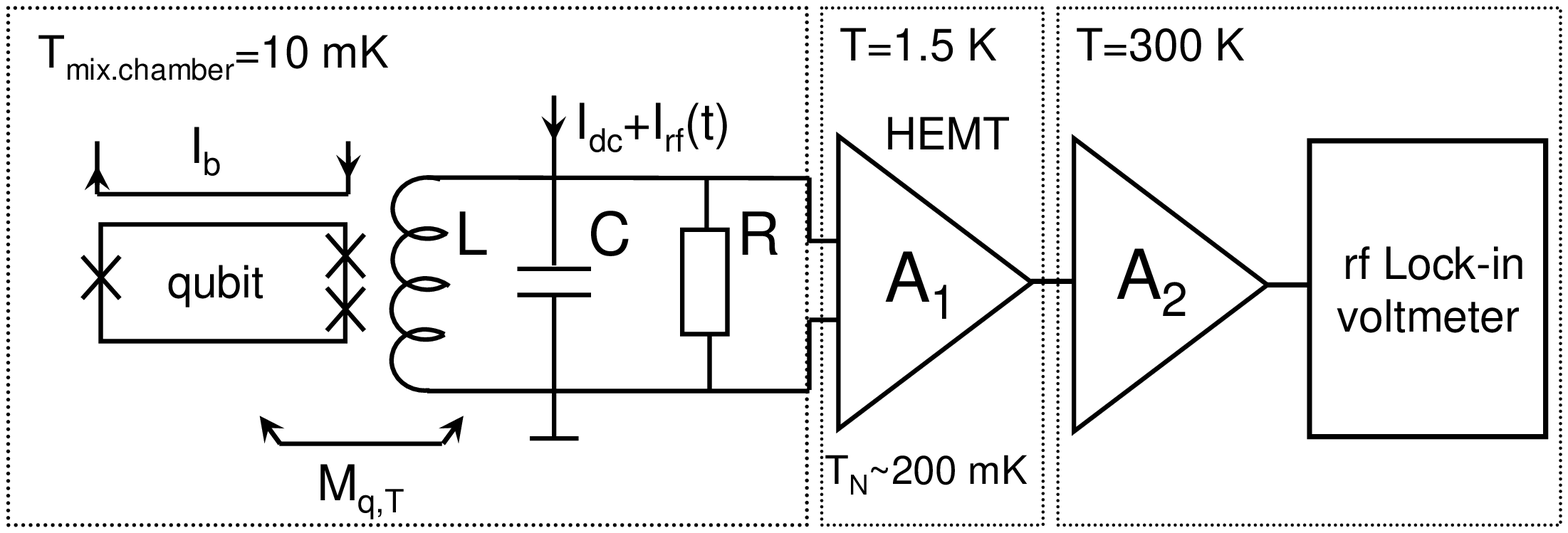} \caption{Scheme of
a parametric transducer inductively coupled to a superconducting
flux qubit. The $rf$ voltage across the tank  is amplified by a
cooled HEMT amplifier thermally linked to a
1~K~pot.\cite{Oukhansky02} After room temperature amplification the
signal is detected by an $rf$~lock-in voltmeter. Both the amplitude
and phase of the $rf$ voltage are measured as a function of the
external magnetic flux applied to the qubit produced by the currents
$I_{dc}$ and $I_{b}$ through a coil and a wire, respectively.}
\label{fig:transducer}
\end{figure}

Parametric transducers have been shown to be very
sensitive instruments, that can overcome the standard quantum
limit.\cite{Braginsky} The precision of the measurement of small
changes of the dielectric susceptibility by a capacity transducer is
of the order of $10^{-10}$. In addition, a parametric transducer can
work in a regime that satisfies the criteria of quantum
nondemolition (QND) measurements.  Usually, an electromagnetic
auto-oscillator is used as a key element of a parametric transducer
since the frequency can be measured with a very high accuracy. The
scheme of a parametric transducer is shown in
Fig.\ref{fig:transducer}, it contain a high quality $LC$ resonator
connected to an amplifier.\cite{Oukhansky02}
The resonant frequency of the $LC$ circuit depends on both the
inductance $L$ and the capacitance $C$ by the relation
$\omega_r=1/\sqrt{LC}$. In our experiments, typically
$\omega_r/2\pi\sim 30$~MHz. This satisfies $\omega_r\ll\omega_q$,
where $\omega_q$ is the transition frequency between the ground and
first excited energy level of the qubit. Thus, the magnetic
susceptibility of the qubit placed in a resonator can be measured
from the shift of the resonance frequency. It can be easily
shown\cite{Braginsky} that the tangent of the phase shift $\theta$
between the voltage across the tank and driving current  is
proportional to the real part of $ac$ susceptibility $\chi'$
\begin{equation}
\tan \theta = -k^2Q\chi' \label{Eq:theta}
\end{equation}
where $0<k<1$ is the coupling coefficient between the resonator and
sample.  The ideas behind a parametric transducer were also used in
the design of an $rf$-SQUID by Silver and Zimmermann.\cite{Silver67}
It was shown theoretically that an $rf$-SQUID can achieve the
quantum limit.\cite{Buhrman} Therefore, the parametric transducer is
a suitable readout device for superconducting flux qubits.

The magnetic susceptibility of the superconducting flux qubit
is\cite{Smirnov}
\begin{equation}
    \chi'= L_qI_q^2\frac{\Delta^2}{(\Delta^2+\varepsilon^2)^{3/2}}
\tanh\left(\frac{\sqrt{\Delta^2+\varepsilon^2}}{T}\right)
\label{Eq:chi}
\end{equation}
where $\Delta$  is the tunneling amplitude, $L_q$ is the inductance
of the flux qubit, $I_q$ is the persistent current in the qubit, $T$
is the temperature, and $\varepsilon=\Phi_0I_q f$ is the bias of the
qubit, where $f$  is the deviation from degeneracy defined in terms
of internal magnetic flux in the qubit as $f=\Phi_i/\Phi_0-0.5$. By
using Eqs.~(\ref{Eq:theta}),(\ref{Eq:chi}) the persistent current
and the tunneling amplitude can be determined experimentally by
measuring the resonator phase as a function of the external magnetic
flux $\Phi_e$.\cite{Grajcar03} The function $\chi'(f)$
(Eq.~\ref{Eq:chi}) has a simple form, and it is easily seen that
$\chi'(f)$ exhibits a peak at the degeneracy point $f=0$. If the
temperature $T\ll\Delta$, the explicit equations for the persistent
current and the tunneling amplitude can be readily derived
\begin{eqnarray}
    I_q=\frac{\Phi_0}{L_q}\frac{\chi'_a f_{FWHM}}{2\sqrt{2^{2/3}-1}} \\
    \Delta=\Phi_0I_q\frac{ f_{FWHM}}{2\sqrt{2^{2/3}-1}}
\end{eqnarray}
where $\chi'_a$ and $f_\mathrm{FWHM}$ are the peak amplitude and the
full width at half maximum (FWHM), respectively.

\begin{figure}
\centering \includegraphics[width=8cm]{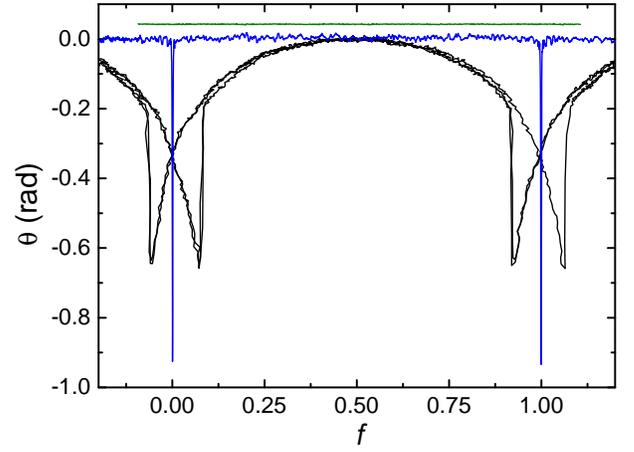} \caption{The phase
shift $\theta$ between the bias current $I_{rf}$ and the $rf$
voltage of the parametric transducer inductively coupled to the
superconducting flux qubit as a function of the internal magnetic
flux in the qubit. The curve with hysteretic behavior (black curve)
corresponds to the 'qubits' with a large ratio $g=E_J/E_C\sim 10^3$
(classical regime). The straight line (vertically shifted for
clarity) and the non-hysteretic line correspond to qubits with
$g\approx 60$, and $\alpha=0.9$ and $\alpha=0.8$, respectively. }
\label{Fig:crossover}
\end{figure}

Here we would like to point out that the measurement by means of a
parametric transducer is a quantum nondemolition measurement,
because the qubit is staying in its ground state the entire time of
the measurement, as the resonant frequency of the resonator
$\omega_r$ is much lower than the transition frequency $\omega_q$.
The output signal of the parametric transducer contains information
about the amplitude of the persistent current, but holds no
information about the phase of the rapidly oscillating persistent
current. A parametric transducer cannot even distinguish whether the
current flows clockwise or counterclockwise. This can be directly
seen for 'qubits' in the classical regime (see the hysteretic curve
in Fig.~\ref{Fig:crossover}).
Exactly at the degeneracy point $f=0$, the two branches of the
hysteretic curves corresponding to current flowing clockwise and
counterclockwise cross, i.e. the transducer gives the same signal.
The reason for this is that the operator probed by the inductive
transducer is $\sigma_x$ as we show below. In this sense, such a
readout is complemental to the SQUID readout which measures
$\sigma_z$ ($\sigma_x$ and $\sigma_z$ are Pauli matrices) to make
the analogy with Stern-Gerlach apparatus complete.\cite{SG} More
formally, the Hamiltonian of a qubit-resonator system at the
degeneracy point $f=0$ can be written in the form\cite{Haken}
\begin{equation}
    H=H_r+H_q+H_{int}=\hbar\omega_rb^\dagger b+\Delta\sigma_x+\gamma (b^\dagger+b)\sigma_z
\label{Eq:H}
\end{equation}
where $b^\dagger, b$ are creation and annihilation operators,
respectively, of the photon field in the resonator,
$\gamma=k\sqrt{\hbar\omega_rL_q}I_q$ is the coupling energy
between the resonator and qubit, and $\sigma_x$ and $\sigma_z$ are
Pauli matrices of the natural basis of the qubit (i.e. the two
eigenstates of operator $\sigma_z$ correspond to the currents
flowing clockwise and counterclockwise). After unitary
transformation
\begin{equation}
    U_1=\frac{1}{\sqrt{2}}
\left(\begin{array}{rr}
1  & 1\\
1  & -1
\end{array}\right)\;,
\end{equation}
the Hamiltonian (\ref{Eq:H}) takes the form
\begin{equation}
    U_1HU_1^\dagger=\hbar\omega_r b^\dagger b+\Delta\sigma'_z+\gamma(b^\dagger\sigma'_-+b\sigma'_+)
\end{equation}
where
\begin{equation}
    \sigma'_+=\left(\begin{array}{rr}
0  & 1\\
0  & 0
\end{array}\right)\;,
\sigma'_-=\left(\begin{array}{rr}
0  & 0\\
1  & 0
\end{array}\right)\;
\end{equation}
are spin-flip operators and $\sigma'_z$ is the Pauli matrix in the
eigenbasis of the qubit at the degeneracy point. Following the
approach in Ref.~\onlinecite{Blais04}, after a second unitary
transformation
\begin{equation}
    U_2=\exp\left(\frac{\gamma}{2\Delta}(b\sigma'_+-b^\dagger\sigma'_-)\right),
\end{equation}
and by expanding to second order in $\gamma/\Delta$, the
transformed Hamiltonian $H'=U_2U_1HU_1^\dagger U_2^\dagger$ is

\begin{equation}
    \frac{H'}{\hbar\omega_r}=\left(1-k^2\frac{W_q}{\Delta} \sigma'_z\right)b^\dagger b+
    \left(\frac{\Delta}{\hbar\omega_r}-\frac{k^2}{2}\frac{W_q}{\Delta}\right)\sigma'_z
\label{Eq:Hres}
\end{equation}
where $W_q=L_qI_q^2/2$ is the magnetic energy of the qubit.
Experimentally, a shift of the resonant frequency of the resonator,
which depends on the qubit state, is measured. This shift is
determined by the first term of Eq.~\ref{Eq:Hres}, i.e. the measured
observable is $\sigma'_z$, and one can readily find that the
sufficient condition for QND measurement $[\sigma'_z,H']=0$ is
satisfied.\cite{Braginsky} Provided that the coupling between
resonator and qubit is small $\sigma'_z$ corresponds to $\sigma_x$
in the original basis. This means that the resonator measures the
observable $\sigma_x$ at the degeneracy point in contrast with the
SQUID which measures $\sigma_z$. Let us point out additional
difference between resonator and SQUID measurement. The SQUID
measurement makes a projection of the spin into the $z$-axis, i.e.
at the degeneracy point the qubit is localized in one of the
classical states after the measurement, and the SQUID gives a signal
corresponding to this state. This measurement is non-QND since the
SQUID is coupled directly to the oscillating
variable.\cite{Averin02} On the other hand, the resonator gives no
signal if the qubit is in the eigenstate of the operator $\sigma_z$,
i.e. the resonator does not perform a measurement, and therefore,
does not disturb the qubit. Such a readout method has a clear
advantage in the case of adiabatic quantum computing. The qubit
remains in its ground state also after the measurement, i.e. the
measurement of one qubit does not spoil the result of the adiabatic
evolution. However, it should be noted that this statement is valid
only if the amplitude of the circulating current in the resonator is
small enough to avoid Landau-Zener transitions. Nevertheless, as we
have shown theoretically in Ref.~\onlinecite{Greenberg} and
experimentally demonstrated in section~\ref{Sec:exp}, the noise
temperature of the cooled amplifier\cite{Oukhansky02} enables one to
fulfill this condition.

The readout procedure could be as follows; let us suppose that the
qubit is in the state $|1\rangle$ (i.e.  $f>0$, see
Fig.~\ref{Fig:en_diag}). If the internal magnetic flux in the qubit
is changing towards zero, then the qubit is moving through its
degeneracy point ($f=0$) where two classical energy levels cross
(dashed lines in Fig.~\ref{Fig:en_diag}). At this point the qubit is
in the superposition of the states $|0\rangle$ and $|1\rangle$ where
the magnetic susceptibility of the qubit changes rapidly. Thus, the
inductive transducer gives a considerable signal. On the other hand,
if the qubit is in the state $|0\rangle$ (i.e. $f<0$), one should
increase the external magnetic flux to move the qubit to the
degeneracy point. If we do not know the state of the qubit at
$f=f_0$ we can sweep the external magnetic flux in order to change
the internal magnetic flux in the qubit around this point and from
the response of the parametric transducer we can determine whether
the qubit was in state $|0\rangle$ or $|1\rangle$ (signal is
observed for $f>f_0$ or $f<f_0$, respectively). In the next section
we will show numerically that the qubits can be readout one after
another while staying all the time in the ground state of the
system.

\section{Adiabatic evolution}
\subsection{Theory}
The idea of quantum  computation by adiabatic evolution is very
simple but, surprisingly, was discovered only
recently.\cite{Farhi00,Kaminsky04} It is based on the fact that,
in practice, it is very difficult to find a ground state of
certain Hamiltonians. Such a task belongs to the set of
non-polynomial time (NP) problems. On the other hand, some
Hamiltonians have a trivial ground state which is easy to find.
Let us assume that the Hamiltonian of $N$ qubits $H(p)$ can be
externally controlled by the parameter $p$ and that its ground
state is separated from the first excited state by the energy gap
$g(p)=E_1(p)-E_0(p)$ ($E_0,E_1$ are the two lowest eigenvalues of
the Hamiltonian $H(p)$). Provided that the ground state of the
initial Hamiltonian $H_I=H(p=0)$ can be easily found,  we can
construct it and then change the parameter $p$ slowly from $p=0$
to $p=1$.  If we do it sufficiently slowly, i.e. in a time
$\tau\gg\hbar\varepsilon_{max}/g_{min}^2$ where the
$\varepsilon_{max}\sim\max E_1(p)-\min E_0(p)$ and $g_{min}=\min
g(p)$,  the ground state of $H_I$ is evolved to the state which is
with high probability the ground state of $H_P=H(p=1)$. Thus, we
have prepared the system of the qubits in the ground state of
Hamiltonian $H_P$ and they can be read-out. As a matter of fact
the system is in the ground state of the Hamiltonian $H(p)$ during
the whole adiabatic evolution, i.e. the system is immune against
dephasing and relaxation. Here we should emphasize that the
adiabatic evolution of the Hamiltonian is crucial in speeding up
considerably the finding of the ground state of the Hamiltonian
$H_P$. One could suggest that it is enough to wait a while and the
system would relax itself into the ground state. However, a
Hamiltonian which encodes a NP problem exhibits a lot of local
minima and the physical system needs an exponentially long time
(as a function of the  number of qubits) to find its global
minimum. As an example, one can consider Ising model of $N$
antiferromagnetically coupled magnetic moments. It is well known
that such a system can be highly frustrated. The task of finding
the minimum of the Ising Hamiltonian is equivalent to the
optimization MAXCUT problem which belongs to a NP-complete
problem.\cite{Garey76} Thus, it seems that NP problems cannot be
solved in polynomial time on either digital or analog classical
computers. Theoretically it was shown\cite{Farhi02} that an
adiabatic quantum algorithm can find the global minimum of some
functions in polynomial time whereas a classical simulated
annealing algorithm requires exponential time. The crucial
condition for adiabatic quantum evolution is the existence of an
energy gap between the ground and upper levels. This is the key
difference between classical and quantum systems, thereby enabling
an enormous speed up of adiabatic quantum algorithms over
classical ones. The size of the energy gap limits the speed of
adiabatic quantum evolution as we will show experimentally in the
next section.

\begin{figure}
\centering \includegraphics[width=6cm]{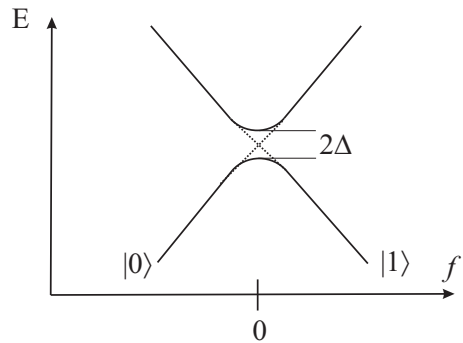} \caption{Quantum
energy levels of the qubit as a function of normalized internal
magnetic flux$f=\Phi_i/\Phi_0-0.5$. For $f$ much less or greater
than zero, the qubit is in the state $|0\rangle$ or
$|1\rangle$,respectively. The dashed lines correspond to the
classical potential minima.} \label{Fig:en_diag}
\end{figure}

Adiabatic evolution can be demonstrated on a single qubit. Following the
original paper by Farhi et al.,\cite{Farhi00} we start from the initial
Hamiltonian at $t=0$
\begin{equation}
    H_I=\Delta\sigma_x\;.
\end{equation}
Then we adiabatically evolve from $H_I$ to the problem
Hamiltonian $H_P$ in time $\tau$
\begin{equation}
H_P=\varepsilon(\tau)\sigma_z\;.
\end{equation}
This scheme can be implemented for a superconducting flux qubit.
Near the degeneracy point $f=0$,
the qubit can be described by
the Hamiltonian
\begin{equation}\label{eq01}
  H(t)= \varepsilon(t)\sigma_{\!z} +\Delta\sigma_{\!x}\;.
\end{equation}
At a bias $\varepsilon=0$, the two lowest levels of the qubit
anticross (Fig.~\ref{Fig:en_diag}), with a gap of~$2\Delta$.
By increasing $\varepsilon$ slowly enough, the qubit will
adiabatically transform from the superposition state
$(|0\rangle+|1\rangle)/\sqrt{2}$ to $|1\rangle$, but remains in the
ground state. For $|\varepsilon(\tau)|\gg\Delta$,  $\Delta$
diminishes and the Hamiltonian takes the form
\begin{equation}
    H(\tau)= \varepsilon(\tau)\sigma_{\!z}\;.
\end{equation}

However, if the bias  changes in time $\varepsilon(t)=\lambda t$,
the qubit can 'jump' from the ground state $|g\rangle$ to the
excited state $|e\rangle$ with probability
$P_{LZ}=\exp(-\pi\Delta^2/\hbar\lambda)$. This process, known as a
Landau-Zener transition,\cite{LZ} would violate adiabatic evolution
and, therefore, should be avoided. This puts constraints on the
characteristic time $\tau$ of the adiabatic evolution which can
globally be estimated as: $\tau\gg\hbar E_J/\Delta^2$. Consequently,
$\tau$ can be considerably shorter if we take into account that a
Landau-Zener transition takes place only in the $\Delta$ vicinity of
the anti-crossing point. Thus, $\varepsilon(t)$ can be changed
quickly except in the region close to anti-crossing point. For such
a local adiabatic evolution the requirement for $\tau$ reads
$\tau\gg\hbar/\Delta$. Note that only this condition leads to a
quadratic speed-up of the adiabatic evolution version of Grover's
algorithm.\cite{Roland02} A measurement by a parametric transducer
provides the unique possibility of controlling the speed of an
adiabatic evolution. The smaller the energy gap is the larger is the
signal from the transducer (see Eq.\ref{Eq:theta},\ref{Eq:chi}).
This signal can be used as feedback for $\varepsilon(t)$ sweeping so
that the condition for adiabatic evolution can be satisfied locally
for an unknown ground state of the system.

The tunnel splitting $2\Delta$ is very sensitive to the Josephson
and Coulomb energy of the junctions. It can be finely tuned by
reducing the size of one junction in the superconducting loop, while
leaving the two others unchanged. If the ratio between the area of
the small and large junction is $\alpha$ $(\alpha<1)$, $\Delta$ can
be roughly estimated\cite{Greenberg,Orlando}
\begin{multline}
    \Delta=\frac{E_\mathrm{J}}{\pi}\sqrt{\frac{2\alpha{-}1}{\alpha g}}\\
  \times\exp\!\left[\sqrt{\frac{g(2\alpha{+}1)}{\alpha}}\left(\arccos\frac{1}{2\alpha}
  -\sqrt{4\alpha^2{-}1}\right)\!\right],
\end{multline}
where $g=E_J/E_C$. By changing the parameters $\alpha$ and $g$,  one
obtains a crossover from the classical, through the Landau-Zener, to
the adiabatic regime.

\subsection{Experiment}
\label{Sec:exp} 
\begin{figure}
\centering \includegraphics[width=8cm]{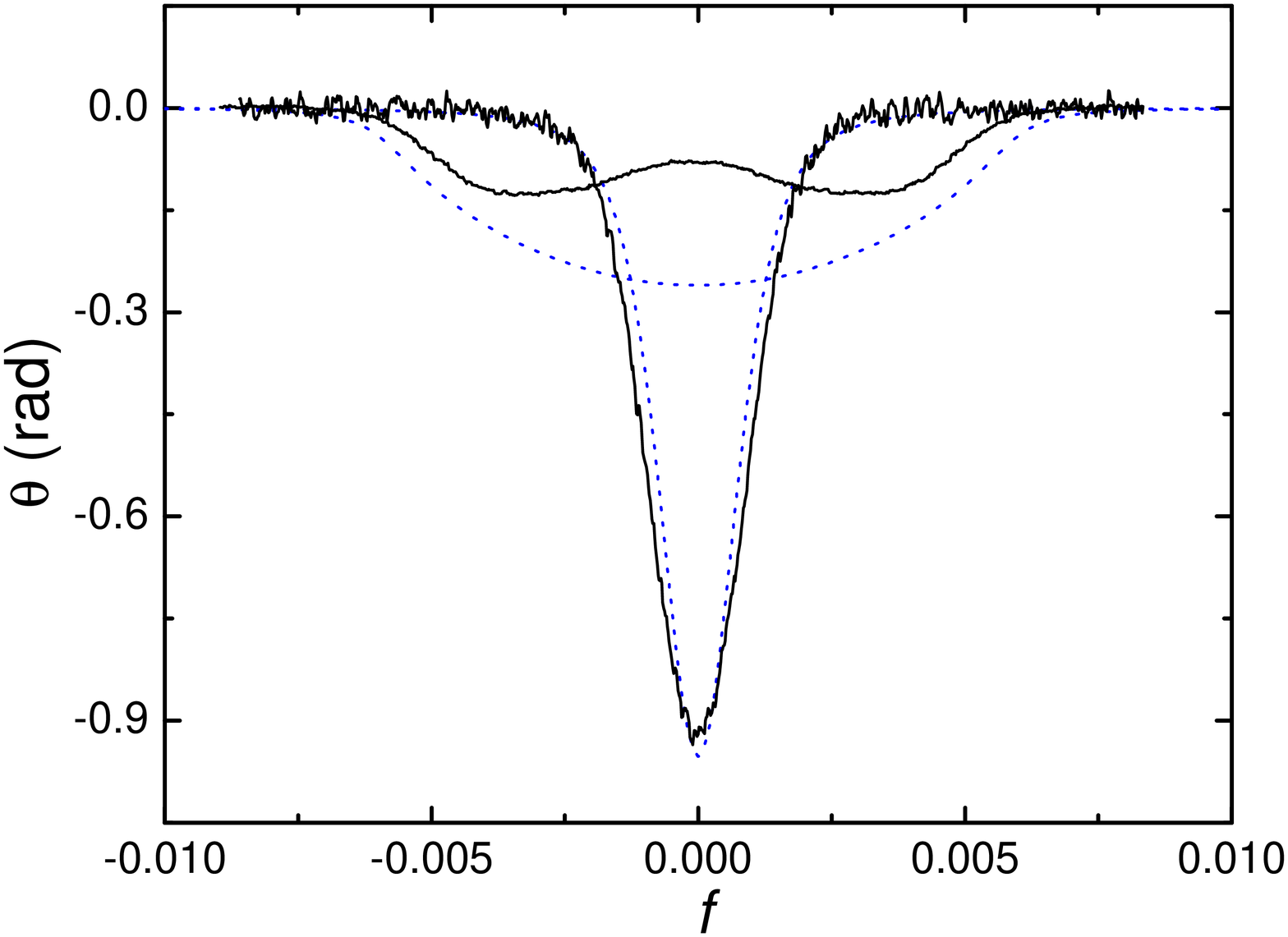} \caption{The phase
shift $\theta$ between the bias current $I_{rf}$ and the $rf$
voltage of the parametric transducer as a function of the normalized
internal magnetic flux for small $V_{rf}\approx0.5 \mu$V (lower
curves) and large $V_{rf}\approx5 \mu$V (upper curves) $rf$
voltages. The resonant frequency of the parametric transducer was 32
MHz. The discrepancy between experimental (solid line) and
theoretical (dotted line) curves for the large amplitude $rf$
voltage is caused by Landau-Zener transitions.} \label{Fig:lz_ad}
\end{figure}

In order to demonstrate the crossover from the
classical to quantum regime we have prepared three qubits with
different parameters $\alpha$ and $g$. The qubits were placed inside
pancake niobium coil made by using electron-beam lithography on
oxidized Si substrates. The typical linewidth and the distance
between 20-30 coil windings are 1-2  $\mu m$. The coils
self-inductances are $L=50-140$~nH. For all experiments reported
here we use an external capacitance $C=470$~pF, therefore the tank
resonance frequency is $19.6-32.8$~MHz with quality factors
Q=700-1700. The 3JJ qubit structure was fabricated out of Al in the
middle of the coil by a conventional shadow evaporation technique.
The critical current was determined, by measuring an rf-SQUID
prepared on the same chip, as $I_C=250-400$~nA. The qubit's loop
area was 90 $\mu m^2$, with $L_q=40$~pH. The typical coupling
coefficient between the coil and qubit is  $1-2 \times 10^{-2}$. In
Fig.\ref{Fig:crossover} the typical response of the inductive
transducer is shown for three values of the parameters $\alpha$ and
$g$, which correspond to three different regimes: classical,
Landau-Zener and adiabatic. In classical regime the signal from
parametric transducer is proportionate to the first derivative of
the Josephson current with respect to internal magnetic
flux.\cite{APL} Close to degeneracy point there are two classical
states corresponding to the currents flowing clockwise and
counterclockwise (hysteretic behavior).  For $g=60$ and $\alpha=0.9$
the qubit is in an intermediate regime where both tunneling between
two classical states and Landau-Zener transitions are not
negligible. There is still no visible dip in the phase
characteristic but the losses caused by Landau-Zener transitions
decrease the quality factor of the resonant circuit and,
consequently, the amplitude of the $rf$ voltage.\cite{Izmalkov} By
keeping $g$ constant, but decreasing the size of the third junction
from $\alpha=0.9$ to $\alpha=0.8$, the tunnel splitting $2\Delta$
increases and Landau-Zener transitions are suppressed. As a result,
a shift of the resonance frequency of the parametric transducer
leads to huge dips in the  $\theta$ vs $f$ curves (adiabatic
regime). Nevertheless, if the voltage amplitude across the
parametric transducer is increased high enough, the Landau-Zener
transitions suppress the dip again. Under this condition, a
discrepancy between experimental and theoretical curves, calculated
within the adiabatic approach is observed (Fig.~\ref{Fig:lz_ad}).
Thus, we have observed the crossover from the classical, through the
Landau-Zener, to the adiabatic regime of a superconducting flux
qubit by decreasing the size of the Josephson junctions. Our
experimental results show that the idea of adiabatic quantum
computing can be demonstrated on a system of superconducting flux
qubits. A reasonable and primarily feasible design is shown in the
next section.

\section{Implementation of the MAXCUT problem for a set of inductively coupled
superconducting qubits}
The MAXCUT problem is a part of the NP-complete
problems. Mathematically, in order to solve the MAXCUT problem, one should find
the maximum of the payoff function\cite{Steffen03}
\begin{equation}
    P(|s\rangle)=\sum_iw_is_i+\sum_{i,j}s_i(1-s_j)w_{i,j}
\label{eq:Ps}
\end{equation}
where $w_{ij}$, $w_i$ are the parameters of the problem and
$s_i=0,1$ are components of the vector $|s\rangle$.
The problem can be encoded into a Hamiltonian $H$ of $N$ inductively
coupled superconducting qubits
\begin{equation}
    H=\sum_{i=1}^N\varepsilon_i(f_i)\sigma_{z,i}+\sum_{i=1}^N\Delta_i\sigma_{x,i}+
    \sum_{i<j}^NJ_{i,j}\sigma_{z,i}\sigma_{z,j}
\label{Eq:H3q}
\end{equation}
where $\sigma_x$ and $\sigma_z$ are Pauli matrices,
$\varepsilon_i(f_i)$ is the energy bias of the $i$-th qubit, and
$J_{i,j}$ is the coupling energy between the $i$-th and $j$-th
qubit. The eigenvector $|s\rangle$, corresponding to the ground
state of the Hamiltonian $H$, is the solution of the payoff function
$P(|s\rangle)$ if (a) $\Delta_i\ll J_{i,j}\ \forall i,j$, and (b)
$\varepsilon_i=-w_i/2$, and $J_{i,j}=w_{i,j}/2$.

 \begin{figure}[tbp]
\centering \includegraphics[width=6cm]{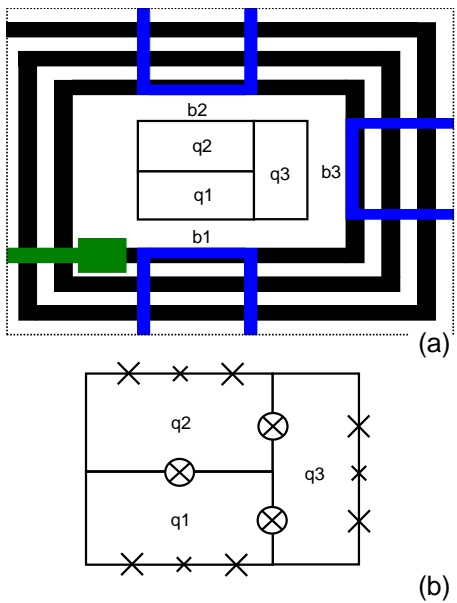}
\caption{Three-Qubit design for MAXCUT problem. (a) Three
superconducting flux qubits are placed in superconducting coil. The
qubits can be biased independently by $dc$ bias wires $b1, b2, b3$.
(b) The qubits are coupled through common Josephson junction marked
by circles. The coupling energy\cite{Levitov01}
$J_{i,j}=(M_{i,j}+\Phi_0/2\pi I_c) I_{qi}I_{qj}$, where $I_c$ is
Josephson critical current of the common junction.} \label{fig1}
\end{figure}

For superconducting qubits, the initial Hamiltonian $H_I$ can be
easily constructed by taking into account that  $J_{i,j}=0$ and
$\Delta_i=0$ if $f_i=-0.5$, i.e.
\begin{equation}
    H_I=\sum_{i=1}^N\varepsilon_i(-0.5)\sigma_{z,i}
\end{equation}
The ground state of $H_I$ is trivial, $|0\rangle$. By changing the
bias of individual qubits adiabatically to $\varepsilon_i=-w_i/2$,
the $H_I$ is transformed to $H$. (The coefficients $w_{i,j}$ are set
by design and they are determined by coupling energies between
qubits.) $H$ encodes the payoff function $P(|s\rangle)$ completely
if $\Delta_i=0$. Unfortunately, we cannot switch off the tunnel
splitting $\Delta_i$ in superconducting qubits, but it is not
absolutely necessary if $J_{i,j}\gg\Delta_i$. Nevertheless, we will
show that by making use of a parametric transducer,\cite{Grajcar03}
one can obtain the answer even if $J_{i,j} \gtrsim \Delta_i$.
Moreover, the qubit states can be readout while staying in the
ground state of the system.

The most simple but still reasonable example of the adiabatic
quantum optimization algorithm MAXCUT can be implemented by three
coupled superconducting flux qubits ($N=3$). The coupling between
the qubits can be realized by means of a common Josephson
junction\cite{Levitov01} shared between two qubits (see
Fig.~\ref{fig1}). This enable to increase the coupling energy over
pure magnetic one. The coupling energy $\approx 0.3$~K has been
measured recently.\cite{Grajcar04} Thus, for the present design we
have chosen the interaction energies between the qubits to be
$J_{1,2}=J_{2,3}=J_{1,3}=0.3$~K, the persistent currents are
$I_{p1}=I_{p3}=350$~nA, $I_{p2}=420$~nA and tunneling matrix
elements are $\Delta_1=\Delta_2=\Delta_3=96$~mK. By choosing
appropriate values for $\varepsilon_i$ it is possible to realize the
situation that the system exhibits both a local and a global
minimum. We have chosen the following parameters
$\varepsilon_1(0.006)=0.315$~K, $\varepsilon_2(0.004)=0.252$~K, and
$\varepsilon_3(0.01)=0.525$~K. The energy of the ground state for
various vectors $|s\rangle$ is shown in Table~\ref{tab:en}. In the
state $|101\rangle$ the system is in the global minimum. Note that
for $|110\rangle$ the system exhibits a local minimum, that is,
there is no way to decrease the energy of the system by flipping the
persistent current in one qubit only. Thus, the system can stay in
the state $|110\rangle$ for an exponentially long time at low
temperatures. In our design the lowest 'energy' barrier which the
system sees from the local minimum is higher than 0.5~K. This could
lead to a wrong answer, unless the Hamiltonian transform is carried
out adiabatically.

The qubits' state can be readout by an inductive transducer as was
described above. The internal magnetic flux of the individual qubits
can be changed by a current through the wires placed nearby each of
them. In such a configuration, all three qubits can be readout by
making use of one transducer only. Nevertheless, the idea should be
checked since qubits interact and $\Delta$ is nonzero. The three
qubit Hamiltonian can be solved numerically. In the following
section we simulate the readout of a parametric transducer
inductively coupled to three superconducting flux qubits.

\begin{table}
    \center
\begin{tabular}[c]{c||c|c|c|c|c|c|c|c}
    $|s\rangle$ & 000 & 010 & 011 & 001 & 101 & 111 & 110 & 100 \\
    \hline
    $E$ (K)   &  1.992  &  0.288 &  -0.342  & 0.162 & -0.889 & -0.192 & -0.762 & -0.258
\end{tabular}
\caption{Energy  of the system for various vectors.
$J_{1,2}=J_{2,3}=J_{1,3}=0.3$~K. $\varepsilon_1=0.315$~K,
$\varepsilon_2=0.252$~K, and $\varepsilon_3=0.525$~K. }
\label{tab:en}
\end{table}

\subsection{Numerical simulation}
\begin{figure}[tbp]
\centering \includegraphics[width=7cm]{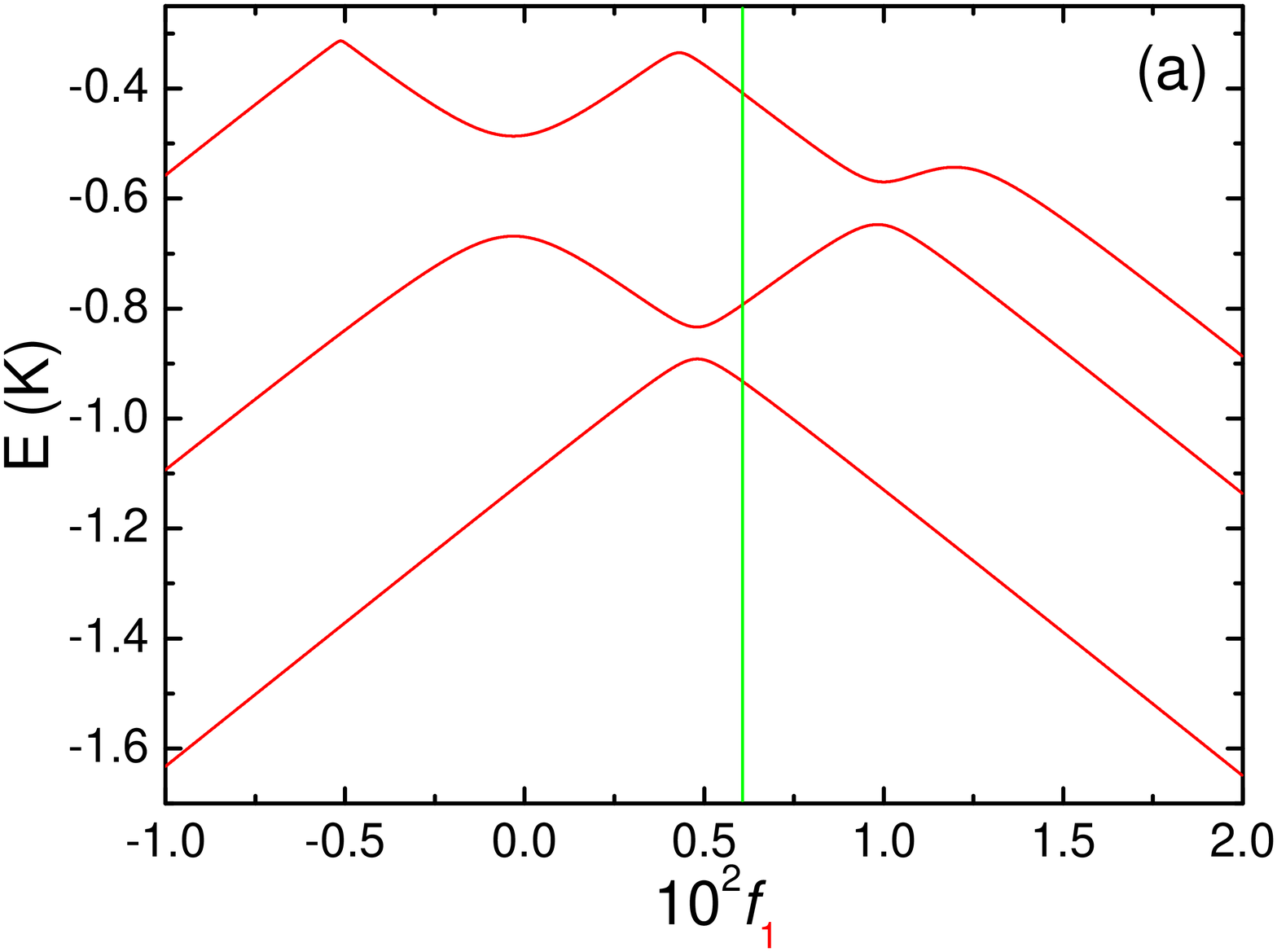}\\
\centering \includegraphics[width=7cm]{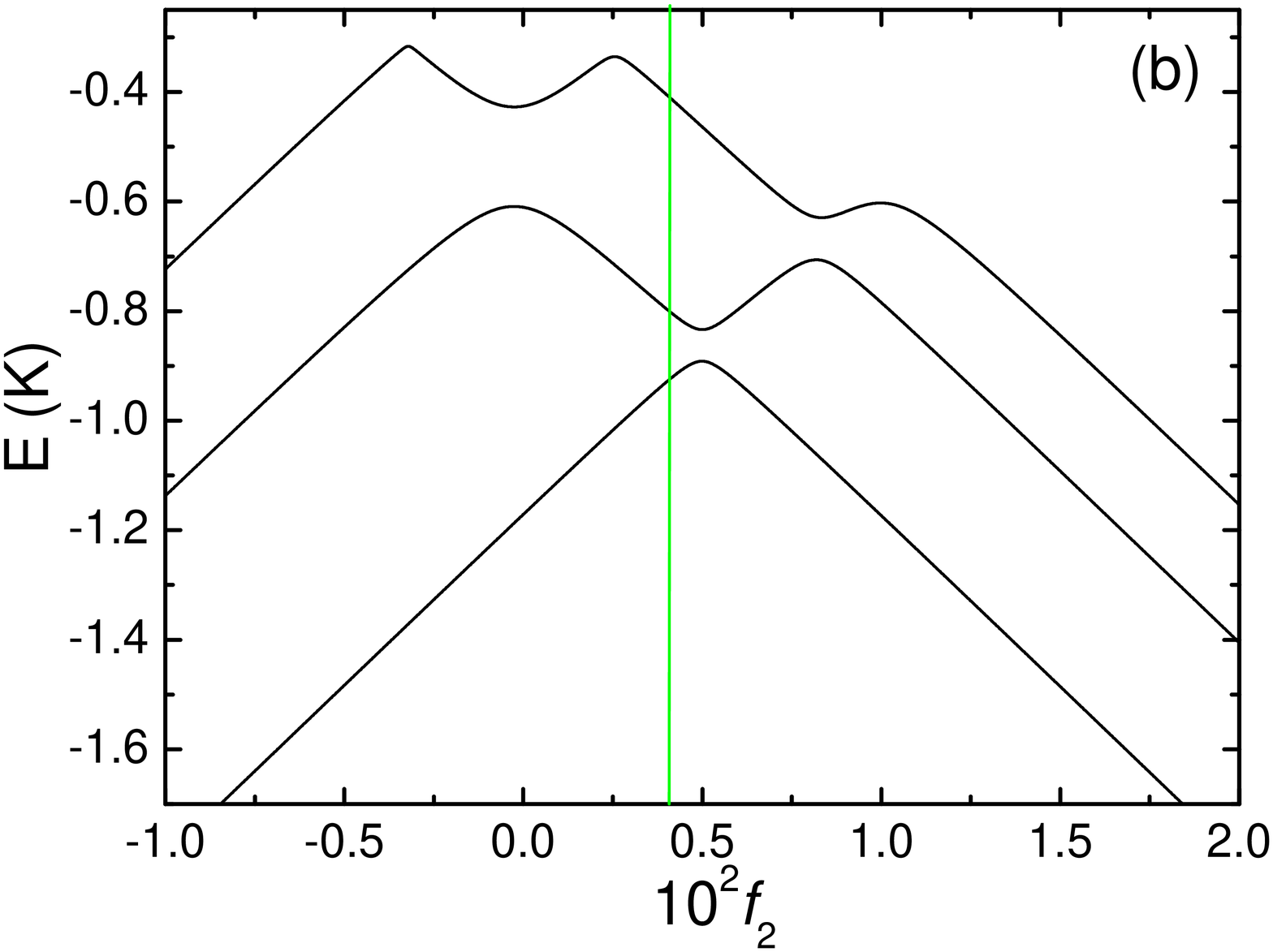}\\
\centering \includegraphics[width=7cm]{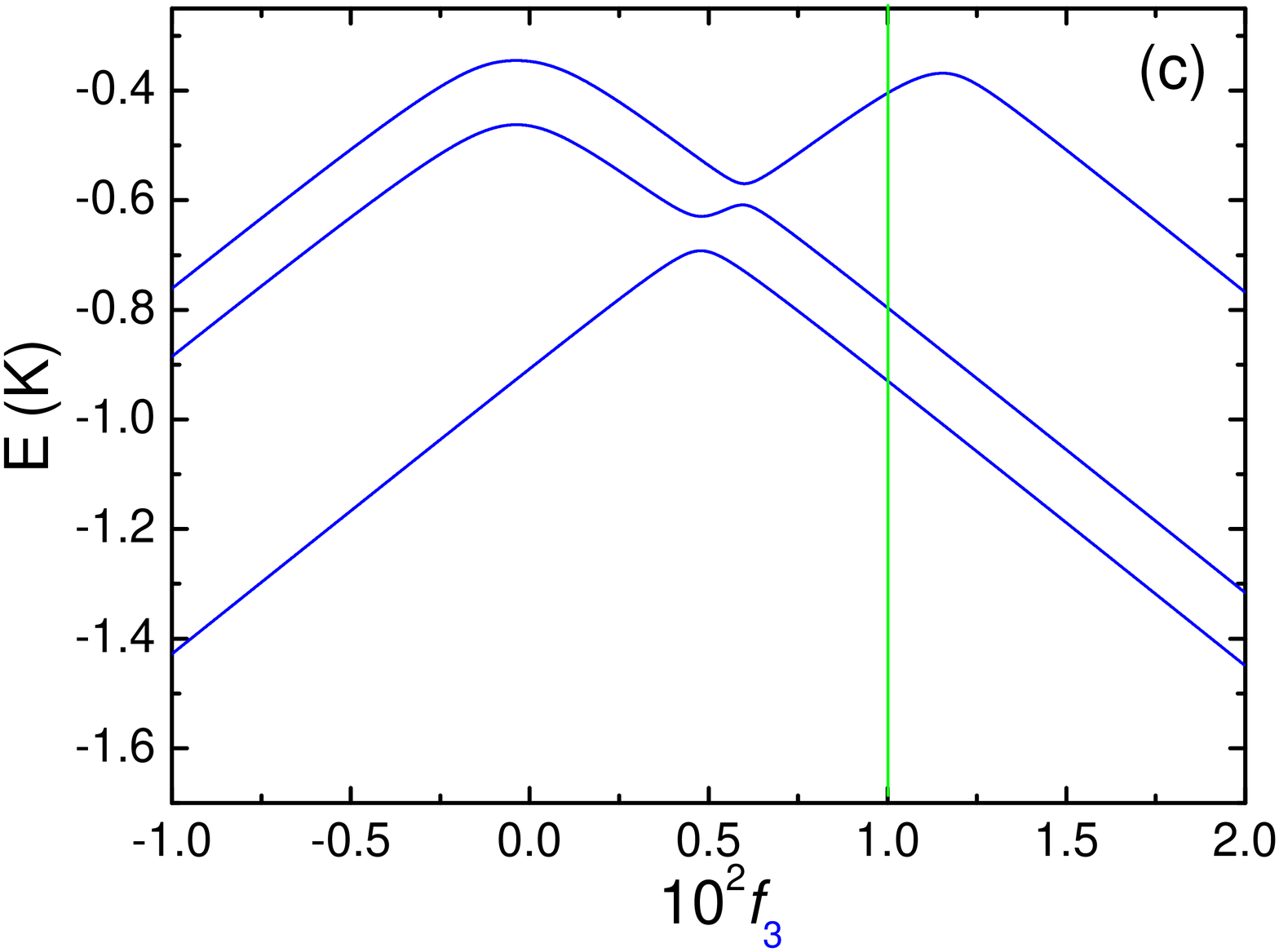} \caption{First
three energy levels of the three qubit system during readout.
Readout of the qubit starts at point $f^p_1=0.006$, $f^p_2=0.004$,
and $f^p_3= 0.01$ then its bias is changed adiabaticaly and
separately  through qubit 1 (a), 2 (b), and 3 (c) while keeping it
fixed in the others. At the points  with a large curvature of the
ground level the parametric transducer gives a considerable response
(see Fig.~\ref{fig:d2E}). If this point is on the left (right) side
of the point corresponding to the problem Hamiltonian (marked by
green vertical line) the qubit is (or better to say would be if
$\Delta=0$) in the state $|0\rangle$ ($|1\rangle$).}
\label{fig:En_lev}
\end{figure}

\begin{figure}[tbp]
\centering \includegraphics[width=8cm]{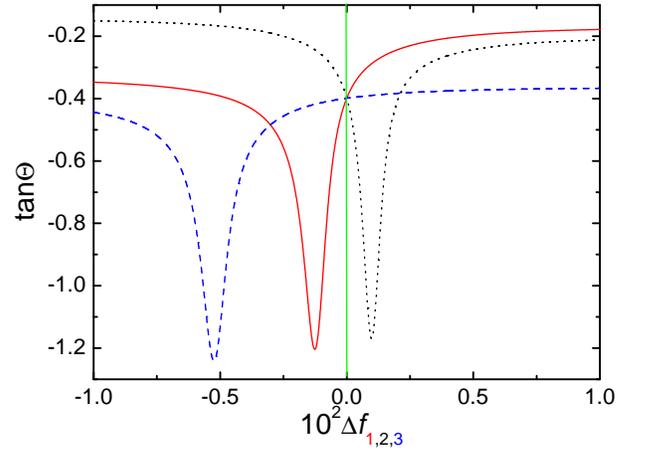}
\caption{The phase shift between the voltage and bias current of the
parametric transducer with respect to $\Delta f=f-f^p$. Readout of
the qubit starts at point $f^p_1=0.006$, $f^p_2=0.004$, and $f^p_3=
0.01$ then its internal magnetic flux is swept adiabaticaly around
this point. The red (solid), black (dotted) and blue (dashed) lines
correspond to bias flux change in qubit 1,2 and 3, respectively.
From the position of the dips we find that the state $|101\rangle$
corresponds to the global minimum (compare with
Table~\ref{tab:en}).} \label{fig:d2E}
\end{figure}

\begin{figure}[tbp]
\centering \includegraphics[width=8cm]{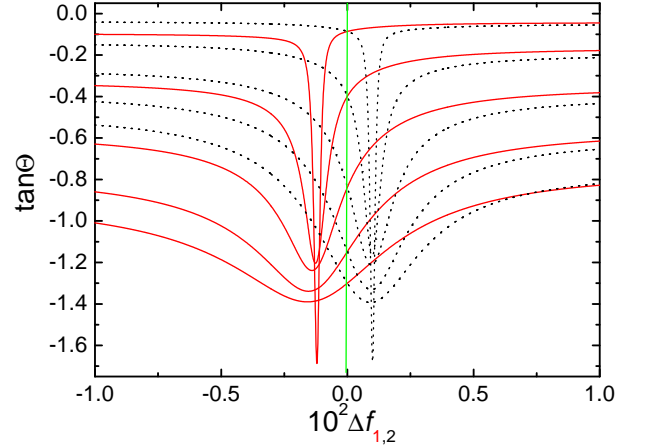}
\caption{The phase shift between voltage and bias current of the
inductive transducer with respect to $\Delta f=f-f^p$ for various
values of $\Delta_i$ ($\Delta_i$ is taken to be the same for all
qubits). Readout of the qubit starts at point $f^p_1=0.006$,
$f^p_2=0.004$, and $f^p_3= 0.01$ then the internal magnetic flux is
swept adiabaticaly around this point. The red solid line and black
dotted line correspond to the qubits 1 and 2, respectively. From the
upper to lower curve (at $\Delta f=0.01$) $\Delta_i$ takes the
values 0.048, 0.096, 0.144, 0.192, and 0.240~K.}
\label{fig:threshold}
\end{figure}

The Hamiltonian (\ref{Eq:H3q}) was solved numerically and the energy
levels of the Hamiltonian (\ref{Eq:H3q}) as a function of $f_i$ are
shown in Fig.~\ref{fig:En_lev}. We have used the same parameters as
those used in our design. We have also calculated the response of
the parametric transducer using the formula \cite{Izmalkov04,
Smirnov03}
\begin{equation}\label{eq_tan_theta}
  \tan \theta = - 2Q \sum_\nu\frac{R_{0\nu}}{ E_\nu - E_0 } .
\end{equation}
where $E_\nu - E_0$ is the distance between ground and upper energy
levels, and
\begin{eqnarray}\label{eq_tan_theta_R}
R_{0\nu} &=& \left(\sum_{i=1}^Nk_i\sqrt{L_{qi}}I_{qi} \langle 0|
\sigma_{z}^{(i)}|\nu\rangle\right) \nonumber\\
&&{}\times \left(\sum_{j=1}^Nk_j\sqrt{L_{qj}}I_{qj} \langle
\nu|\sigma_{z}^{(j)}|0\rangle\right)
\end{eqnarray}
are the real matrix elements. Here $k_i$ is the coupling coefficient
between the $i$-th qubit and resonator, and $L_{qi}$ is the qubit's
inductance. For $Q=1000$, $L=81$~nH, $L_{qi}= 40$~pH, and
$k_i=0.036$ the results are shown in Fig.~\ref{fig:d2E}. From these
figures it is apparent that the qubits' states can be determined by
a parametric transducer. We have also tried to find the threshold
for $\Delta_i$ below which the state of the qubit cannot be
distinguished. As a criterion the existence of the distinguishable
dips on the experimental curves can be chosen. From
Fig.~\ref{fig:threshold} one can see that the positions of the dips
do not change as $\Delta$ increases and they can be distinguished
for relatively large values of $\Delta$. Thus, the parametric
transducer readout delivers the right solution of the problem.

\section{Conclusions}
Experimentally, we have demonstrated  the principle of adiabatic
quantum evolution in a single qubit. Theoretically, we have shown
that three inductively coupled superconducting flux qubits placed
in a superconducting coil can be used  to demonstrate the
adiabatic quantum algorithm MAXCUT which belongs to the set of
NP-complete problems. A three qubit design has been proposed and
simulated numerically.

Note added in proof: Recently, Lupascu et al. [A. Lupascu, C. J.
M. Verwijs, R. N. Schouten, C. J. P. M. Harmans, and J. E. Mooij,
Phys. Rev. Lett. 93, 177006 (2004)] proposed a similar readout
method which enables to measure the observable  $\sigma_z$ in a
nondestructive way.

\section{Acknowledgement}
The authors thank D-Wave Systems Inc. for financial support and
M.H.S.~Amin, D.V.~Averin, A.~Blais, Alec Maassen van den Brink,
Ya.S.~Greenberg, H.E.~Hoenig, H.-G.~Meyer, Yu.A.~Pashkin,
A.~Shnirman, F.K.~Wilhelm, and A.M.~Zagoskin for fruitful
discussions. M.G. wants to acknowledge partial support by Grant
Nos. VEGA 1/2011/05 and APVT-51-016604.




\begin{thebibliography}{99}
\bibitem{SG} The analogy between spin and
the Stern-Gerlach apparatus on the one hand and persistent current
and SQUID on the other is not complete since rotating the SQUID by
$90^\circ$ does not allow the measurement of $\sigma_x$.
\end{thebibliography}
\end{document}